\documentclass[aps,prl,floats,twocolumn,showpacs,superscriptaddress]{revtex4-1}

\usepackage{graphicx}
\usepackage{amssymb}
\usepackage{color}

\begin{document}
\title{Impact of lag information on network inference}
\author{Nicol{\'a}s Rubido}

\affiliation{Universidad de la Rep{\'u}blica, Instituto de F{\'i}sica de la Facultad de Ciencias, Igu{\'a} 4225, 11400 Montevideo, Uruguay }
\author{Cristina Masoller}
\email{cristina.masoller@upc.edu} 
\affiliation{Departament de Fisica, Universitat Politecnica de Catalunya, Rambla Sant Nebridi 22, ES-08222 Terrassa, Barcelona, Spain} 
%
\begin{abstract}
Extracting useful information from data is a fundamental challenge across disciplines as diverse as climate, neuroscience, genetics, and ecology. In the era of ``big data'', data is ubiquitous, but appropriated methods are needed for gaining reliable information from the data. In this work we consider a complex system, composed by interacting units, and aim at inferring which elements influence each other, directly from the observed data. The only assumption about the structure of the system is that it can be modeled by a network composed by a set of $N$ units connected with $L$ un-weighted and un-directed links, however, the structure of the connections is not known. In this situation the inference of the underlying network is usually done by using interdependency measures, computed from the output signals of the units. We show, using experimental data recorded from randomly coupled electronic R{\"o}ssler chaotic oscillators, that the information of the lag times obtained from bivariate cross-correlation analysis can be useful to gain information about the real connectivity of the system. 
\end{abstract}

\maketitle
%
\section{Introduction}
\label{intro}
Network inference involves discovering, from observations, the underlying connectivity between the elements of a complex system. Reliable inference is important because it allows to understand, predict, and control complex behaviours. Main challenges involve the fact that usually one can only observe a single scalar variable (but the evolution of the system depends on other --unobserved variables), during a limited time-interval, with limited resolution, and with considerable measurement noise. In recent years many network inference methods (or network reconstruction) have been proposed \cite{inference_genetics_2006,inference_timme_2007,inference_celso_2011,Nico_2014,Nico_2015,Tirabassi_2015,inference_chaos_2015,inference_kurths_2016,inference_arkady_2016,inference_misha_2017,inference_davidsen_2017,inference_mason_chaos_2017,inference_nat_comm_2017,inference_pre_2017,inference_kurths_2017}, whose success depends, not only on the previous knowledge of the system (e.g., weighted or unweighted interactions, directed or undirected, instantaneous or lagged), but also, on data availability (e.g., hidden nodes or unobserved variables and limited temporal or spatial resolution).

Relevant examples of network inference include brain functional networks and climate networks. Brain functional networks, which have shed light into many neurological conditions, such as Alzheimer, Parkinson or Epilepsy, are inferred from recorded brain signals (magneto-encephalography, MEG, electro-encephalography, EEG, and functional magnetic resonance imaging, fMRI) by correlating different brain regions and linking the ones that exhibit the highest correlations \cite{brain_victor_dante_2005,brain_review_2009,brain_epilepsy_2014}. Similarly, climate networks have shed light into climate phenomena (such as long range tele-connections or atmosphere-ocean interactions) by correlating time-series of climate variables and linking the geographical regions that exhibit significant correlation.
 \cite{climate_havlin_2008,climate_tsonis_2008,climate_donges_2009,climate_klaus_2010,climate_cris_2011}.

When trying to infer a system's connectivity, the statistical similarity of the time series recorded from different units is commonly measured by using bivariate time series analysis, such as cross-correlation or mutual information. Typically, time series are mutually lagged in order to find the maximum of the similarity measure, $S_{ij}$, but the information contained in the set of lag times, $\tau_{ij}$ has {not yet been used to infer the links of the network. In climate network studies, the lag times have been used to infer the directionality of the links; in addition, lag analysis has received attention in the context of financial and ecological data analysis~\cite{Olden_2001,Curme_2015,Damos_2016}}. In this work we investigate if the discovery of the real interactions in a complex network can be improved if these lag times are taken into account. The working assumption is that, when the strength of the coupling is increased, the transition to synchronized behavior occurs \cite{book}. During this transition, the lags between the time-series of nodes which have direct interactions can be smaller than the lags between nodes that are not directly coupled. In other words, if two nodes have a direct link between them, they can synchronize with a lag that is smaller than the lag between nodes that are not directly connected, and this difference can be used for improving network inference.

Here, we analyze under which conditions the lag information can be used to complement the information of the similarity measure for improving the discovery of the existing links (true positives), for decreasing the number of wrongly inferred links (false positives), for improving the detection of non-existing links (true negatives) and for decreasing the number of mistakes due to undetected links (false negatives).

We investigate an experimental dataset from a network of electronic R{\"o}ssler chaotic oscillators \cite{Tirabassi_2015} composed by $N=12$ units, which are randomly coupled with $L=19$ links. The real underlying adjacency matrix, $A_{ij}$, is known, while an inferred matrix, $A^*_{ij}$, is extracted by using bivariate time series analysis. We propose three criteria for classifying links as existing or non-existing and, by comparing the inferred and the known coupling matrices, we discuss the effectiveness of the different criteria for uncovering the real connectivity of the system. We conclude {that when using an OR criterion, namely, one that detects a link when either a small lag or a high similarity value is found, non-existent links are almost always correctly discarded. However, this criterion also discards direct links more than the other two criteria}.

\section{Data}
\label{sec:Data}
The data was described in \cite{Tirabassi_2015}. It is generated from $N=12$ R{\"o}ssler electronic oscillators randomly coupled with $L=19$ links. The coupling between units $i$ and $j$ is $kA_{ij}$, where $k$ is the coupling strength and $A_{ij}$ is the adjacency matrix [$A_{ij}=1$ {if} the oscillators $i$ and $j$ are coupled and $A_{ij}=0$ if they are not]. The dataset contains the time series recorded for 31 values of the coupling strength (the minimum is $k=0$ and the maximum is $k=0.15$).  Each time series has 30000 data points. In order to reduce the effects of noise, each time series is divided into non-overlapping segments of length $T$. To avoid transient effects the first segment is disregarded, and the following segments are used for the analysis. The results presented are obtained with $T=5000$, so we have five segments for computing the mean values and the error bar of the measures described in the following section.

\section{Methods}
\label{sec:methods}
The lagged cross-correlation is used to quantify the similarity between the time series recorded from nodes $i$ and $j$, $x_i(t)$ and $x_j(t)$, with $t\in [1,T]$ and $i,j\in [1,N]$. Specifically, each time series is first normalized to zero-mean and unit variance. Then, we calculate the Pearson coefficient, 
\begin{equation}
C_{ij}(\tau)=\frac{1}{T-\tau_{max}} \left | \sum_{t=1}^{T-\tau_{max}} x_i(t)\,x_j(t+\tau) \right |,
 \label{eq_CC}
\end{equation}
varying $\tau$ in the interval $[0,\tau_{max}]$ with $\tau_{max}=T/5$~\cite{Tirabassi_2015}.  We define the lag, $\tau_{ij}$, between nodes $i$ and $j$ as the value of $\tau$ that maximizes $C_{ij}(\tau)$, and we define the correlation strength, $S_{ij}$, as {$S_{ij} \equiv C_{ij}(\tau_{ij})=\max_\tau C_{ij}(\tau)$}.

Next, we test whether the information contained in the matrices $\tau_{ij}$ and $S_{ij}$ is useful for inferring the existing links. We define two thresholds, one for the lags, $\tau_{th}$, and one for the correlation strengths, $S_{th}$, and use the following criteria for classifying the links between {the} existing and the non-existing {ones.}

\begin{enumerate}
\item {\textbf{SIM}:} Only the similarity measure (CC) is used to infer the links. The link between $i$ and $j$ exists ($A^*_{ij}=1$) if $S_{ij}>S_{th}$, else, the link does not exist ($A^*_{ij}=0$).
\\
\item {\textbf{AND}}: The link between $i$ and $j$ exists ($A^*_{ij}=1$) if  $\tau_{ij}<\tau_{th}$ and $S_{ij}>S_{th}$, else, the link does not exist ($A^*_{ij}=0$).
\\
\item {\textbf{OR}}: The link between $i$ and $j$ exists ($A^*_{ij}=1$) if  $\tau_{ij}<\tau_{th}$ or $S_{ij}>S_{th}$, else, the link does not exist ($A^*_{ij}=0$).
\\
\end{enumerate}
For these criteria, the thresholds $S_{th}$ and $\tau_{th}$ are chosen such that they return a number of links as close as possible to the (known) number of existing links.

To quantify the efficiency of these criteria for uncovering the real connectivity of the network we use the following measures

\begin{itemize}
\item True negatives (TN): number of non-existing links which are correctly classified as not existing, relative to the number of non-existing links;
\item False negatives (FN): number of existing links which are incorrectly classified as not existing, relative to the number of existing links;
\item True positives (TP): number of existing links which are correctly classified as existing, relative to the number of existing links;
\item False positives (FP): number of non-existing links which are incorrectly classified as existing, relative to the number of non-existing links.
\end{itemize}

We also quantify the global success of the inference method by calculating the total wrongly predicted existent, FP, and non-existent, FN, links relative to the total number of links:
\begin{equation}
\Delta = \frac{2}{N\,(N - 1)}\sum_i \sum_{j> i} |A_{ij}-A^*_{ij}| = \frac{FN + FP}{N\,(N - 1)/2}.
 \label{eq_Delta}
\end{equation}

Finally, since both $\tau_{ij}$ and $S_{ij}$ depend on the length of the segment, $T$, of the time series, and of the maximum lag, $\tau_{max}$, we analyze if the results are robust with respect to the choice of these parameters.

\section{Results}
\label{sec:results}
\begin{figure}[tbp]
\center \resizebox{0.7\columnwidth}{!}{\includegraphics{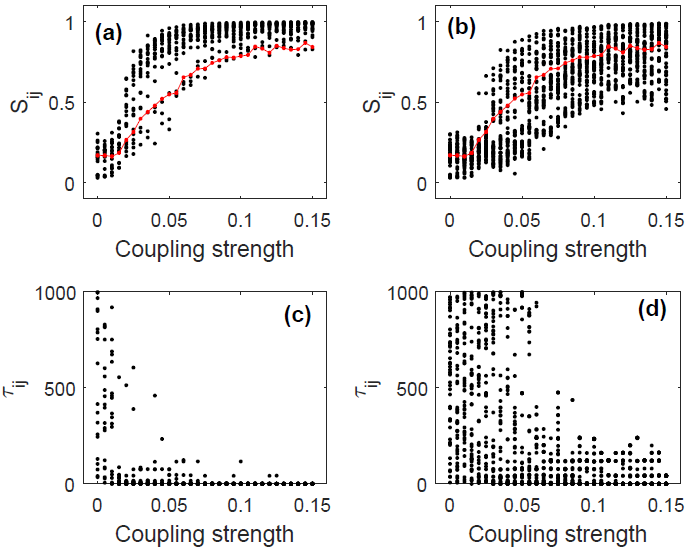} }
  \caption{Analysis of experimental data ($N = 12$ R{\"o}ssler electronic oscillators coupled in a random network): comparison of the values of the similarity measure, $S_{ij}$, and lags {, $\tau_{ij}$,} of the existing links [panels {\bf (a)} and {\bf (c)}, respectively] and of the non-existing links [panels {\bf (b)} and {\bf (d)}, respectively]. It is observed that the variation of $S_{ij}$ and $\tau_{ij}$ with the coupling tends to be different: for the existing links, $S_{ij}$ increases faster {(and $\tau_{ij}$ decreases faster) than for} the non-existing links. In panels (a), (b) the solid lines indicate the value of the similarity measure averaged over all (existing and non-existing) links.}%
 \label{fig0}
\end{figure}

In Fig.~\ref{fig0} we {show} the values of the similarity measure, $S_{ij}$, {[see Eq.~(\ref{eq_CC})]} and the corresponding lag, $\tau_{ij}$ separating the links that exist ($A_{ij}=1$, left column {panels}) and the links that do not exist ($A_{ij}=0$, right column {panels}). We clearly observe a different variation as the coupling strength is increased: for the existing links, $S_{ij}$ tends to increase faster in comparison with the non-existing links, and the opposite happens with $\tau_{ij}$. 

\begin{figure}[tbp]
\center\resizebox{0.9\columnwidth}{!}{\includegraphics{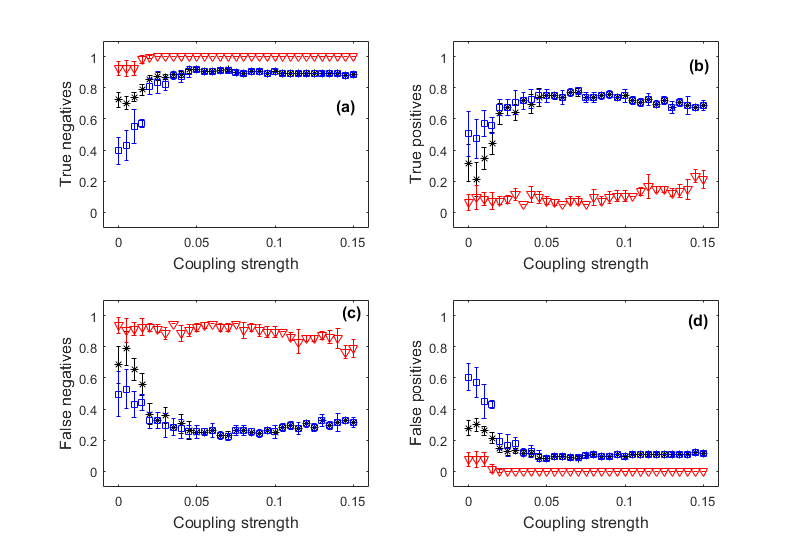} }
\caption{Quantitative comparison of the inference of the network obtained with the three criteria described in Sec.~\ref{sec:methods}: SIM (*), AND (squares) and OR (triangles). {The analysis is performed in six non-superposed segments of $T=5000$ data points each. In each segment a network is inferred and then the true positives (TPs), false positives (FPs), true negatives (TNs) and false negatives (FNs) are calculated. The symbols and the error bars indicate the corresponding mean values and standard deviations, computed from the TP, FP, TN and FN values obtained in the six segments}.}
\label{fig:1}       
\end{figure}

\begin{figure}[tbp]
\center\resizebox{0.4\columnwidth}{!}{\includegraphics{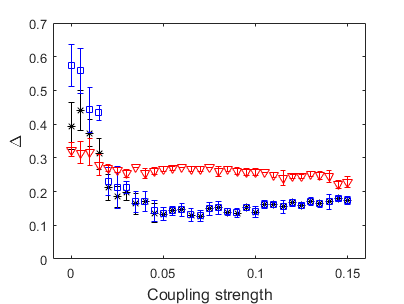} }
\caption{{Comparison of the total number of mistakes, as defined in Eq.~(\ref{eq_Delta}), for the three inference criteria: SIM (*), AND (squares) and OR (triangles).}}
\label{fig:3}       
\end{figure}

Next, we show how including the lag information into the bivariate analysis for the network inference can be useful. Figure~\ref{fig:1} displays the different types of errors that are made when applying the criteria described in Sec.~\ref{sec:methods}. We see that the SIM and AND only differ for weak coupling, but as the coupling increases, the number of correctly inferred links and mistakes made are the same for the two criteria. At weak coupling, adding the lag information (AND) improves the detection of the existing links (true positives), at the cost of also improving the detection of not-existing links (false positives). When considering the total mistakes, as defined in Eq.~(\ref{eq_Delta}), we can see in Fig.~\ref{fig:3} that the AND criteria produces, at low coupling, more mistakes than the SIM criteria. Consequently, in this system the lag information is helpful, at low coupling, to decrease particular types of mistakes of the inference process. However, using similarity values alone is better if the main goal is to minimize the total number of mistakes{, i.e., the sum of wrongly predicted existing, FP, and non-existing, FN, links}.

Interestingly, the OR criteria gives very different results: avoids the false positives at the cost of giving a large number of false negatives. This is due to the fact that, regardless of the coupling strength, many $\tau_{ij}$ values are small for both, existing and non-existing links (see Fig.~\ref{fig0}). 

\begin{figure}[tbp]
\center\resizebox{0.8\columnwidth}{!}{\includegraphics{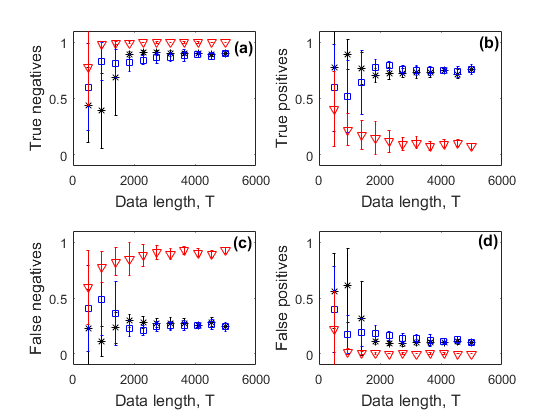} }
\caption{{Robustness of the results obtained with the three inference criteria, with respect to the length, $T$, of the data segment for coupling strength $k=0.1$. For $k=0$ there is no effect of $T$ (not shown).}}
\label{fig:T}       
\end{figure}

\begin{figure}[tbp]
\center\resizebox{0.8\columnwidth}{!}{\includegraphics{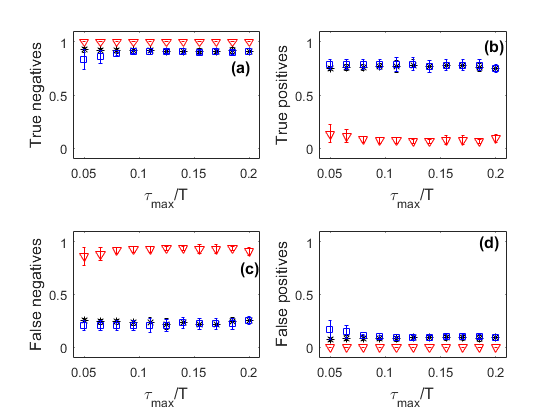} }
\caption{{Robustness of the results obtained with the three inference criteria, with respect to the length, $\tau_{max}$, of the interval where we search for the maximum of the cross-correlation, normalized to the length of the data segment. The coupling strength is $k=0.1$; for $k=0$ there is no effect of $\tau_{max}$ (not shown).}}
\label{fig:tau}       
\end{figure}

The analysis of how relevant the particular choice of $T$ and $\tau_{max}$ is, indicates that these results are robust. In Fig.~\ref{fig:T} we consider a given coupling strength ($k=0.1$) and change the length of the segment, $T$, of the time series, while in Fig.~\ref{fig:tau} we change the interval of lag values where we search for the maximum of the cross correlation. In both cases we see that for $T$ and $\tau_{max}$ large enough, the number of {FPs, TPs, FNs, and TNs are independent of} the choice of $T$ and $\tau_{max}$.

\section{Conclusions and discussion}
\label{sec:conclusions}

We have explored the possibility of using the lag information from pairwise cross-correlation for improving the inference of the connectivity of a system composed by interacting units. We have limited the study to a simple situation: the number of units and the number of links between pairs of units are known, and the links either exist or do not exist (i.e., they are undirected and unweighted).

We have used data recorded from by 12 electronic R{\"o}ssler chaotic oscillators, coupled in a random network that has 19 links. We have found that lag information can be useful for reducing certain types of mistakes, but it does not improve network inference when all the mistakes are added up. {In particular, we find that we can decrease the number of false positive detections and detect all true negatives in a robust way -- independent of the coupling strength value -- when using an OR criterion. Regarding the total number of errors, guided by Fig. 3 we conclude that, in our system, for weak coupling the OR criterion is the best option, while for intermediate coupling SIM performs the same as AND and both outperform OR. If the coupling is large enough to synchronize the system, it is not possible, using these criteria, to infer the network.}

{In general, a} drawback of {including} the lag information is that one needs to select two inference thresholds, $S_{th}$, {for the similarity values,} and $\tau_{th}${, for the lags}. Here we used the simplest approach: we varied them simultaneously [$S_{th}$ was increased linearly from $\min(S_{ij})$ while $\tau_{th}$ was decreased linearly from $\max(\tau_{ij})$], and the pair of values $S_{th}$ and $\tau_{th}$ that returned a number of links closer to the known number of existing links were used for the inference. {This approach requires a minimal knowledge from the network, namely, its link density -- number of links and network size.} However, a set of $S_{th}$ and $\tau_{th}$ values that returned the target number of links was not always found. A possible way to improve the inference method is by considering two independent thresholds; however, an important consideration is the shape of the distributions the $S_{ij}$ and $\tau_{ij}$ values: if they are bimodal or long-tailed, there might not be any combination of thresholds that returns the target number of links, because a small variation of one of the thresholds might result in either too many or too few links being classified as existent. 

{In a realistic situation the number of existing links is unknown. Therefore, choosing a set of thresholds $S_{th}$ and $\tau_{th}$ that return a pre-defined number of links is not an appropriated inference strategy. In this situation a} possible alternative for using lag information for network inference is by taking into account how the lags and the similarity measures vary with the coupling strength. Here we have not used the fact that when the link between units $i$ and $j$ indeed exists, $S_{ij}$ ($\tau_{ij}$) tends to increase (decrease) with the coupling faster than when the link does not exist. However, classifying links according to the variation of $S_{ij}$ and $\tau_{ij}$ with the coupling, increases the data requirements, as the values of $S_{ij}$ and $\tau_{ij}$ will need to be compared for different coupling conditions. In addition, in systems with non-instantaneous interactions (i.e., coupling delays) or in systems where the units display periodic behavior, the lag information will not be useful for network inference because in such systems the units can synchronize with lags between them which do not have a clear relation with the underlying interactions.

\section*{Acknowledgments}
C. M. acknowledges partial support from Spanish MINECO (FIS2015-66503-C3-2-P) and from the program ICREA ACADEMIA of Generalitat de Catalunya. NR acknowledges the support from the {4th} CSIC MIA {2017} {(id 194)} program, Uruguay. Both authors gratefully acknowledge R. Sevilla-Escoboza and J. M. Buld{\'u} for the permission to analyse the experimental data sets in \cite{Tirabassi_2015} and \cite{data_sets}.


\end{document}